\documentclass[10pt, conference, compsocconf]{IEEEtran}
\usepackage{verbatim}

\ifCLASSINFOpdf
\else
\fi
\usepackage[ruled,vlined]{algorithm2e}

\ifCLASSINFOpdf
  \usepackage[pdftex]{graphicx}
   \graphicspath{{figs/}}
  \DeclareGraphicsExtensions{.pdf,.jpeg,.png}
\else
\fi

\usepackage{multicol}
\begin{document}
%
\title{IoT based Platform as a Service for Provisioning of Concurrent Applications 
}



%
\author{\IEEEauthorblockN{ Deepak kumar Aggarwal \IEEEauthorrefmark{1}}
Rajni Aron \IEEEauthorrefmark{2},
\IEEEauthorblockA{\IEEEauthorrefmark{1}CICS\\
Montreal, Canada
}
\IEEEauthorblockA{\IEEEauthorrefmark{2}Lovely Professional University, India; Machine Intelligence Research Lab\\
 USA
}}


\maketitle

\begin{abstract}

The modern era has seen a speedy growth in the Internet of Things (IoT). As per statistics of 2020, twenty billion devices will be connected to the Internet. This massive increase in Internet connected devices will lead to a lot of efforts to execute critical concurrent applications such fire detection, health care based system, disaster management, high energy physics, automobiles, and medical imaging efficiently. To fasten the emergence of novel applications, this vast infrastructure requires "Platform as a Service(PaaS)" model to leverage IoT things. As a single global standard for all device types and IoT-based application domain is impracticable, we propose an IoT-based Cloud to leverage PaaS model in this paper. This model can host the concurrent application for Wireless Sensor Network (WSN). The proposed model offers the communication interface among processes by uniquely allocating network interface to a particular container.

\end{abstract}

\IEEEpeerreviewmaketitle

\section{Introduction}

Cloud computing has emerged as computing paradigm to solve large-scale problems in the field of science, e-commerce, and industries etc \cite{armbrust2009above}\cite{buyya2009cloud}. Cloud computing works on three types of service models Software as a Service (SaaS), Platform as a Service (PaaS), and Infrastructure as a Service (IaaS). PaaS operates the virtualized infrastructure in order to host and execute concurrent applications. PaaS platform can  play an important role in the development of Internet of Things (IoT) based application. The Internet of Things (IoT) is becoming ubiquitous with sensor nodes getting more intelligent and capable of transmitting their processed data to a cloud \cite{atzori2010internet}\cite{xia2012internet}. Concurrency-based applications play a vital role in the rise of the IoT.

IoT based concurrent application refers to many processes that are connected to the different wireless sensor network. Concurrency is execution of several processes at the same time. A concurrent application is composed of multiple processes, each of which is executing a set of instructions in sequential way [8]. Basically, concurrent application is developed on core foundation of high-performance operating system and storage technology to deliver superior results. Due to important characteristics of concurrent applications such as Parallelism, Availability of services (AoS). Parallelism makes a better use of multiple resources for complex programs in simple multi-core processor architecture e.g., scientific/engineering applications, simulations etc. By AoS, long-running process need not delay short-running ones, e.g., a web server can load an entry page while at the same time processing a complex query is going at the back-end. Task requiring certain preconditions can suspend and wait until the preconditions hold then resume execution transparently. In spite of that, there are so many other challenges for provisioning the critical concurrent applications in a cloud platform such as 1) No Delay 2) Reliable Communication and 3) Dynamic Setting of Parameters.
For provisioning of critical concurrent applications, there should not be delay. Reliable communication means a message sent by one process to another will be received.  
As health care application, traffic control system and forest fire detection application are IoT based concurrent applications having a various process. For the dynamic setting of parameters, one process has to communicate with another process without any delay. A specific requirement for prototyping IoT based concurrent systems is the integration of different techniques such as the following:
\begin{itemize}

\item evaluation of performance, e.g. latency, Throughput
\item simulation of real-time properties
\item evaluation of above mentioned parameters issues
\item transformation of prototype into efficient implementation
\end{itemize}

Traditionally, all applications are hosted on their particular container to provide the isolated environment. If the different application of the same host to communicate with each other, the container has provided the HTTP endpoint. It may cause unacceptable delays for latency sensitive IoT applications such as Disaster management, fire management application as shown in figure 1. It creates a huge impact on the performance of any application. These issues motivate us to work on IoT-based application development on PaaS. There is a need to design IoT-based PaaS architecture for concurrent applications.  To solve the above-mentioned issues, container networking is used. Container networking extends the traditional cloud computing paradigm to provision concurrent application, to avoid the excessive delays. 

The core idea is to realize  the need of  IoT based PaaS framework that provides essential platform services on cloud for IoT solution providers to efficiently deliver and  continuously extend their services. The contribution of this paper is two-fold. First is the design and implementation of an IoT based PaaS architecture. The second contribution of this paper is an approach to provisioning of critical application in the proposed IoT based PaaS architecture. The main aim of this paper is that the proposed architecture handles all issues related to concurrent applications while provisioning.

The rest of the paper is organized as follows: section II described the related work. The high-level architecture of the proposed PaaS with requirements is presented in section III. Mode of operation is also presented. Section IV shows performance evaluation. Section V concludes the paper.

\begin{multicols}{2}
\begin{figure*}[!htb]
	\includegraphics[width=3in]{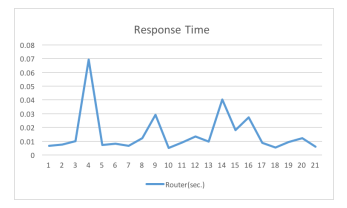}%
	\label{fig:montage}
	\caption{Response Time}
\label{fig:workflows}
\end{figure*}
\end{multicols}
\section{Related Work}
In this section, the existing work for provisioning of concurrent applications in PaaS has been discussed. 

Fei li et al introduced the IoT PaaS architecture, on which IoT solutions can be delivered as virtual verticals by leveraging computing resources and middleware services on cloud. On IoT based PaaS, two types of services related to data are provided to handle real-time events and persisted data respectively. They have focused on domain-specific control applications \cite{li2013efficient}. Luis M. Vaquero et al. focused on to achieve proper PaaS scalability cloud providers must address issues both at container and database level. At container level, a better scalability can be achieved by enabling multi-tenant containers. This imposes strong isolation requirements which maybe not all platforms can achieve by default \cite{vaquero2011dynamically}. 
Lihong Jiang et al. proposed a data storage framework not only enabling efficient storing of massive IoT data, but also integrating both structured and unstructured data. This data storage framework is able to combine and extend multiple databases and Hadoop to store and manage diverse types of data collected by sensors and RFID readers \cite{jiang2014iot}.  
Antonio Celesti et al. introduced a new a new development model for PaaS named "the Distributed Resilient Adaptable Cloud Oriented (DRACO)". They have focused on how DRACO PaaS can be adopted for the development of any kind of specialized PaaS. Basically, The authors have provided a platform for the development of complex algorithms in the Cloud. They have analyzed require a distributed file system able to optimize the concurrent writing operations \cite{celesti2013draco}.  
Vögler, Michael et al \cite{vogler2016scalable} presented LEONORE that supports push-based as well as pull-based deployments. To improve scalability and reduce generated network traffic between cloud and edge infrastructure, they have emphasized on distributed provisioning approach on resource-constrained, heterogeneous edge devices in large-scale IoT deployments.  
we can see that there are a number of papers in which the execution of concurrent application on PaaS is discussed. They have not discussed about delay parameter of critical concurrent applications as this is an important parameter for provisioning in IoT based PaaS.
To the best of our knowledge there are no solutions in literature able to offer both the data-centric and device-centric services at the same time. IoT based PaaS is a new cloud framework able to offer different types of services to its clients. This implies high flexibility and adaptability of the cloud to the specific clients needs.
They have not focused on reduction of latency and due to lack of container networking , time will be consumed by application process more that leads to less throughput.
They have not focused on resource provisioning  of IoT applications in order to effectively utilize resources in cloud data center. This work does not meet the requirement reduce latency, increasing throughput.
\begin{figure*}[!htb]
	\centering
	\includegraphics[width=3in]{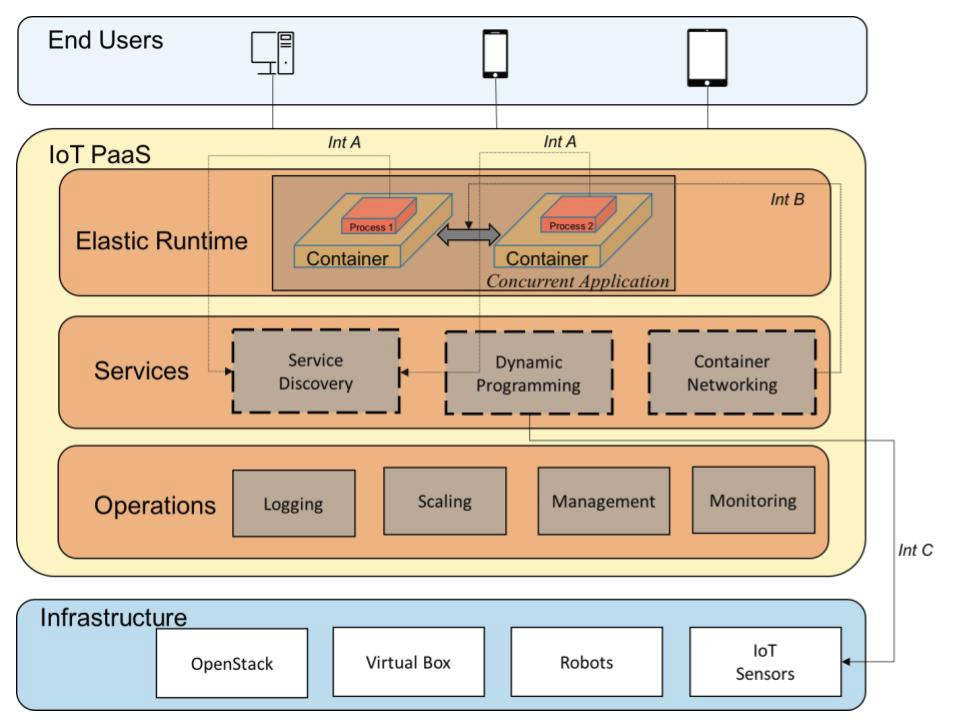}%
	\label{fig:montage}
	\caption{IoT based PaaS: An architecture for Concurrent Application Provisioning}
\label{fig:workflows}
\end{figure*}

\section{IoT based PaaS}

We have proposed a IoT based PaaS architecture where resource providers at IaaS layer give the facility of resource provisioning to the user for optimum results. IoT based PaaS architecture can thus assist organizations in enhancing customer satisfaction for the execution of critical concurrent applications. The requirements of the framework, are as follows:
\subsection{Requirements}
\begin{itemize}
\item Reduce Latency : Critical Concurrent applications should be able to respond quickly. As fire manager application is executed in a shorter time by subdivision into various processes that can be executed concurrently. The architecture must be able to fulfill the latency requirements for real time-based applications. All processes of fire detection application will pass message to neighborhood process quickly so that the early stage precautions could be taken.

\item Scalability: An ideal PaaS should be able to increase scalability for deploying critical concurrent applications.  As the processes of fire manager application are deployed in the different container. Whenever demand increase to control the fire, the architecture should be able to scale a number of resources. VM instance is the resource whenever the alert message will come then number of instances will be initialized. 

\item Dynamic Programming: Different applications may have different requirements for dynamic programming. Classical PaaS should provide the feature dynamic programming. For example, fire detection application hosted on the sensor nodes based on the environment condition. This application may need to change according to the fire affected environment condition. 

\end{itemize}

\subsection{IoT based PaaS Architecture : Mode of Operation}
Figure 1 shows the IoT based PaaS architecture for provisioning of concurrent applications. The architecture executes the requests as follows:

\begin{itemize}
\renewcommand\labelitemi{--}

\item In proposed IoT PaaS architecture, first end user will try to login though any GUI and web based to deploy concurrent applications.
\item  For provisioning of concurrent application in scalable runtime environment, different container will be used. Elastic runtime provides the facility of provisioning in our proposed IoT PaaS Layer.
\item  The processes of concurrent application, which have been hosted on the particular container will register itself with the Service Registry by using interface A.
\item  After the completion of the service registry, container networking will provide the direct route between them by using interface B. 
\item  Dynamic Programming will able to reprogram the IoT sensor devices with the help of interface C. For dynamic programming REST API will be used.
\item  Different operation such as logging, scaling, deployment and management would be done for the execution of application.
\item  On IaaS layer, all sensors and devices will be deployed for execution of application.

\end{itemize}
IoT presently being one of the most pertinent application domains for cloud computing. However, despite the fact that most of the traditional PaaS solutions (e.g. Google App Engine, Microsoft Azure, Cloud Foundry) enable provisioning of IoT applications, but all of them do not support the provisioning of concurrent application. This is due to the absence of direct pipelining between the two application processes. To attain proper execution of the applications with process belonging to an external domain, the proposed PaaS architecture is more efficient than classical PaaS. With container networking, application processes able to communicate directly in the IoT based PaaS architecture \cite{pahl2015containerization}. This complicates more handling the execution flow during the runtime since the application components locations may constantly change.  
The proposed architecture allows provisioning of concurrent application and dynamic programming of sensor devices from the IoT PaaS layer. These above features are not supported by the traditional cloud foundry \cite{cloudfoundry2017} till now.

  \begin{figure*}[!th]
	\centering
	\includegraphics[width=2in]{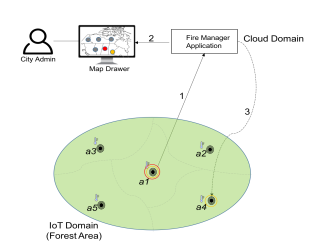}%
	\label{fig:montage}
	\caption{Case Study : Forest Fire Detection}
\label{fig:workflows}
\end{figure*}
 \section{Performance Evaluation}
      
In this section, the implementation of proposed IoT based PaaS for provisioning of critical concurrent application is presented. 
For the development layer implementation, we have used several API like Eureka service discovery and Spring boot framework. Eureka is a REST (Representational State Transfer) based service for locating services for the purpose of load balancing and failover of middle-tier servers. For the execution layer, when processes are executing in the different container, the Eureka service discovery helps to discover the process to whom client has to communicate.  

We have considered early fire detection application. As shown in figure 3, the fire detection application is running on the sensors in the IoT domain and sends the temperature readings to the Gateway constantly. As each sensor associate with particular process of Fire manager application running on the PaaS. The Gateway will stores all the readings to the specified port and forward it to the process of their respective sensors. Suppose temperature exceeds a given threshold for the sensor A, the process A will notify to the process B as shown in the below diagram. Process B will further sends a reprogramming instruction for sensor B via gateway.   
                         
A dynamic reprogramming is required that allows fire manager process to configure the fire detection application when environmental conditions change.  When the application is deployed, the PaaS allocates an application container along with two REST-Based interfaces. One interface is for the VS corresponding to the light sensor and the other for the VS corresponding to the temperature sensor. 

IoT presently being one of the most pertinent application domains for Cloud computing. However, despite the fact that most of the traditional PaaS solutions (e.g. Google App Engine, Microsoft Azure, Cloud Foundry) enable provisioning of IoT applications, but all of them do not support the provisioning of concurrent application in an efficient manner. This is due to the absence of direct pipelining between the two applications processes. To attain proper execution of the applications with process belonging to an external domain, the proposed PaaS architecture is more efficient than classical PaaS. This complicates more handling the execution flow during the runtime since the applications components locations may constantly change.

\section{Conclusions}
In this paper, we introduced IoT based PaaS architecture for cloud environment that enables developing generic and provisioning of critical concurrent applications. The implementation of proposed architecture showed that it enables the provisioning of real time applications in large scale IoT cloud resource pool and at the same time tried to reduce the time.  In future we intend to extend scheduling of these application by considering load balancing constraints.



%

\end{document}